\documentclass[]{aa}
\usepackage{graphicx}
\begin{document}
%---------- page titre-----------------------
\title{On the alignment of Classical T~Tauri stars with the magnetic 
field in the Taurus-Auriga molecular cloud}
%
%\subtitle{Place subtitle here}
%-- liste des auteurs ----
%
\author{
Fran\c cois M\'enard \inst{1}
\and
Gaspard Duch\^ene \inst{1,2}%\fnmsep\thanks{ ca rajoute une footnote!}
}
%-------   
\offprints{Fran\c cois M\'enard}
% ----- liste des adresses ----------
%
\institute{  
  Laboratoire d'Astrophysique, Observatoire de Grenoble, CNRS/UJF
  UMR~5571\\ 414 rue de la Piscine, BP 53, F-38041 Grenoble cedex~9,
  France\\
  \email{menard@obs.ujf-grenoble.fr}
\and  
Department of Physics and Astronomy, UCLA, Los Angeles, CA 90095-1562, 
USA\\
  \email{duchene@astro.ucla.edu} 
}

\date{Received: version 03/02/2004; accepted:}

\abstract{
In this paper we readdress the issue of the alignment of Classical
T~Tauri stars (CTTS) with the magnetic field in the Taurus-Auriga
molecular cloud. Previous studies have claimed that the jet axis of
active young stellar objects (YSO), projected in the plane of the sky,
is aligned preferentially along the projected direction of the local
magnetic field. We re-examine this issue in view of the numerous high
angular resolution images of circumstellar disks and micro-jets now
available. The images show that T~Tauri stars as a group are oriented
randomly with respect to the local magnetic field, contrary to
previous claims. This indicates that the magnetic field may play a
lesser role in the final stages of collapse of an individual
prestellar core than previously envisioned. The current database also
suggests that a subsample of CTTS with resolved disks but without
observations of bright and extended outflows have a tendency
to align perpendicularly to the magnetic field. We discuss the
possibility that this may trace a less favorable topology, e.g.,
quadrupolar, for the magnetic field in the inner disk, resulting in a
weaker collimated outflow.
 }
\authorrunning{M\'enard \& Duch\^ene}
\titlerunning{Alignment of T~Tauri stars in Taurus-Auriga}

\maketitle
% ------------- fin de page titre -----------------
%
\section{Introduction}

The Taurus molecular cloud complex is one of the best studied
star-forming region. However, it is not typical of most other
such regions in that it does not contain massive stars. This
is an advantage for studying low-mass star formation because these
massive stars have a direct and large impact on their
environment, complicating the identification of the processes
responsible for the more quiescent solar-like star
formation. Also important, Taurus is located nearby and
suffers only low extinction. As a consequence, its stellar
population is well determined and the census essentially
complete. The molecular gas distribution is known to be
filamentary (e.g., Schneider \& Elmegreen 1979; Scalo 1990;
Mizuno et al. 1995), and oriented mostly perpendicular to the
projected large scale magnetic field.  These observations
provided early support for the models because, due to the
properties of magnetic forces, collapse is expected to be
easier along magnetic field lines. Except for a few isolated
groups, the young stellar population of the Taurus cloud also
forms filaments, or ``bands'' distributed parallel to the
molecular filaments (Hartmann 2002).

On a smaller scale, Galli \& Shu (1993) showed that the
collapse of an isothermal sphere threaded by straight magnetic
field lines results in a pseudo-disk oriented perpendicular to
the direction of the original magnetic field. If a
centrifugally supported disk forms along the same direction
then one expects, in this simple picture of quasi-static,
magnetically-driven, isolated star formation, that young stars
should have their disks oriented with their major axis
perpendicular to and their jet oriented parallel to the local
magnetic field. Early observations provided support for this
view. Strom et al. (1986) noted during a deep imaging survey
of YSO driving HH objects in Taurus that jets and outflows
have a tendency to align with the local magnetic
field. Similarly, Tamura \& Sato (1989) later argued that
disks were preferentially perpendicular to the local magnetic
field based on a correlation between the linear polarisation
and the magnetic field position angles (PAs).

However, the complete picture of solar-like star formation may prove
once again more complicated than expected. Goodman et al. (1990)
showed that, although the magnetic field may control the large scale
direction of the collapse of molecular clouds in Taurus and Perseus,
it does not do so with an ``iron grip''. If, at large scales, the
cloud structure indeed appears filamentary, elongated perpendicular to
the magnetic field, significant deviations are found between the
direction of the magnetic field and the shape of molecular clouds on
smaller scales. Collapsing cores are expected to be elongated and the
observed average aspect ratio of 2.4 found in the optical by Lee \&
Myers (1999) is comparable with the predictions.  Interestingly, the
cores also show a tendency to be elongated along the filaments (Myers
et al. 1991). However, for statistical reasons, these cores are more
likely prolate than oblate as they would otherwise all need to be
almost perfectly edge-on (Ryden 1996; Curry 2002). This is in apparent
contradiction with the above picture for isolated and quasi-static
star formation (e.g., Hartmann 2002). In particular, it is unclear how
prestellar cores (or filaments) that are elongated perpendicular to
the magnetic field can later produce CTTS with jets now aligned with
the magnetic field.

Here we investigate the influence of the magnetic field on the
process of star formation by re-addressing the issue of the
orientation of CTTS with respect to the magnetic field. We use
recent high-angular resolution images of T~Tauri stars located
in Taurus to find the orientation of the symmetry axis of each
stellar system and compare it to the local magnetic field. In
\S~2 we describe the data sets for the stars and the
orientation of the magnetic field in Taurus. In \S~3 we
present the results, i.e., their relative orientations. In
\S~4 we show that CTTS are oriented randomly with respect to
the magnetic field, the implication of which we discuss in
\S\,5. In \S~6 we discuss the possible impact of the specific
orientation of a source on the properties of the jet it
drives. We summarize our findings in \S~7.

%----------------------------------------
\section{The data}
\subsection{Orientation of T~Tauri stars}
\subsubsection{The database}

In Table~\ref{tab:datatts}, we have compiled a catalogue of
37 classical T Tauri stars (CTTS) in the Taurus molecular cloud
complex for which a spatially resolved jet and/or disk has
been observed. We restrict our study to the zone 4h00 $<
\alpha <$ 5h00 in right ascension and +17\degr$ < \delta <
$+31\degr~ in declination. The final sample contains twenty
six (26) objects in which a jet or outflow is detected and
resolved well enough to estimate its PA. Most of them, fifteen
(15), also have a resolved disk. Additionaly, eleven (11)
objects are tabulated where only a disk is resolved, but
without signs of extended outflow.  In this sample we
considered HH~30 (Burrows et al. 1996) and IRAS~04158+2805
(M\'enard et al. 2004) as normal CTTS, based on spectrocopic
evidence. Their peculiar photometric properties come from
their highly inclined orientation.

Whenever possible we have defined the symmetry axis of a CTTS by the
orientation of its jet. Otherwise, we have assumed that the symmetry
axis of the CTTS lies in the direction perpendicular to the major axis
of its disk. In all cases where both a jet and disk have been resolved
but two (FT~Tau, and DO Tau, but see \S~\ref{sec:remarks}), the PAs of
both structures are perpendicular to one another to $\sim$15\degr\ or
better, as one predicts for example with magneto-hydrodynamical (MHD)
wind models. Since the estimates of the jet and disk orientations are
independent, this result supports the reliability of the PAs presented
in Table~\ref{tab:datatts}.

\begin{table*}[htb]
\caption{Orientation of T~Tauri stars in Taurus-Auriga}
\begin{tabular}{lllccccccc}
\hline
Object & $\alpha (2000)$ & $\delta (2000)$ & P.A. (jet) &
ref. & P.A. (disk) & ref. & P.A. ($\vec{B}$) &
$\rho_{IS}$/$N_{IS}$ & $|\Delta$P.A.$|$ \\
 & & & \degr &  & \degr & & \degr & & \degr \\
(1) & (2)& (3) & (4) & (5) & (6) & (7) & (8) & (9) & (10) \\
\hline
\hline
CW Tau & 04$^{\rm h}$14$^{\rm m}$16\fs95 &
28\degr10\arcmin59\farcs3 & 144$\pm$2 & 1 & -- & -- & 15 &
1.5 / 12 & 51 \\
DD Tau A & 04 18 31.13 & 28 16 30.1 & 125 & 26 & -- & -- & 15 & 0.5 /
5 & 70 \\
CoKu Tau/1 & 04 18 51.54 & 28 20 28.1 & 28 & 11 & 120 & 4
& 15 & 0.5 / 5 & 13 \\
04158+2805 & 04 18 58.2 & 28 12 23 & 165 & 5 & 89 &
5 & 15 & 0.5 / 5 & 30 \\
RY Tau & 04 21 57.34 & 28 26 36.9 & 110 & 14 &
31$\pm$6 & 6 & 18 & 0.5 / 10 & 88 \\ 
T Tau N & 04 21 59.39 & 19 32 06.8 & 90 & 18 & 19 &
19 & 85 & 1.5 / 4 & 5 \\ 
T Tau S & 04 21 59.39 & 19 32 06.1 & 170 & 18 & -- &
-- & 85 & 1.5 / 4 & 85 \\
Haro 6-5B & 04 22 00.89 & 26 57 37.6 & 53 & 7 &
152$\pm$2 & 6 & 22 & 0.5 / 32 & 31 \\ 
FT Tau & 04 23 39.17 & 24 56 15.1 & 29 & 14 &
82$\pm$17 & 8 & 73 & 0.5 / 18 & 44 \\
DG Tau B & 04 27 02.56 & 26 05 30.7 & 111$\pm$8 & 11
& 32 & 4 & 32 & 0.5 / 16 & 79 \\
DF Tau & 04 27 02.80 & 25 42 22.3 & 127 & 26 & -- & -- & 35 & 1.0 / 44
& 88 \\
DG Tau & 04 27 04.67 & 26 06 16.9 & 42 & 9 & 136 &
10 & 30 & 0.5 / 18 & 12 \\
Haro 6-10 & 04 29 23.65 & 24 33 02.0 & 60 & 12 & -- & --
& 74 & 0.5 / 12 & 14 \\
HH 30 & 04 31 37.6 & 18 12 26 & 31 & 23 & 121 & 23 &
77 & 0.5 / 13 & 46 \\
HL Tau & 04 31 38.44 & 18 13 58.9 & 51 & 7 &
125$\pm$10 & 20 & 77 & 0.5 / 13 & 26 \\
XZ Tau & 04 31 40.02 & 18 13 57.8 & 15 & 21 & -- & -- &
77 & 0.5 / 13 & 62 \\
Haro~6-13 & 04 32 15.61 & 24 29 02.3 & 65 & 25 & -- & -- & 
66 & 0.5 / 16 & 1 \\
UZ Tau E & 04 32 42.93 & 25 52 32.1 & 0 & 12 &
86$\pm$2 & 2 & 30 & 1.0 / 16 & 30 \\
GK Tau & 04 33 34.47 & 24 21 07.6 & 20 & 24 & -- & -- & 50 &
0.5 / 19 & 30 \\
DL Tau & 04 33 39.04 & 25 20 38.9 & 145 & 14 &
50$\pm$3 & 2 & 53 & 0.5 / 4 & 88 \\
HN Tau & 04 33 39.32 & 17 51 52.8 & 170 & 12 & -- & -- &
77 & 1.0 / 13 & 87 \\
DO Tau & 04 38 28.57 & 26 10 50.5 & 70$\pm$10 & 12 &
61$\pm$5 & 6 & 56 & 0.5 / 7 & 14 \\
HV Tau C & 04 38 35.29 & 26 10 40.0 & 24 & 14 &
110 & 15 & 56 & 0.5 / 7 & 32 \\
Haro 6-33 & 04 41 38.9 & 25 56 26 & 60 & 14 & -- & -- & 52 &
0.5 / 9 & 8 \\
DP Tau & 04 42 37.66 & 25 15 38.1 & 40 & 16 & -- & -- &
39 & 0.5 / 5 & 1 \\
UY Aur & 04 51 47.36 & 30 47 14.1 & 40 & 12 &
135$\pm$5 & 17 & 65 & 2.0 / 8 & 25 \\
\hline
CY Tau & 04 17 33.73 & 28 20 47.8 & -- & -- &
150$\pm$7 & 2 & 15 & 1.0 / 8 & 45 \\ 
BP Tau & 04 19 15.82 & 29 06 27.9 & -- & -- &
152$\pm$3 & 2 & 175 & 0.5 / 5 & 67 \\ 
IQ Tau & 04 29 51.50 & 26 06 46.5 & -- & -- &
29$\pm$2 & 6 & 26 & 0.5 / 12 & 87 \\
HK Tau B & 04 31 50.55 & 24 24 18.4 & -- & -- & 40 & 13 &
70 & 0.5 / 14 & 60 \\
GG Tau & 04 32 30.27 & 17 31 41.5 & -- & -- &
97$\pm$2 & 22 & 77 & 1.0 / 13 & 70 \\
DM Tau & 04 33 48.70 & 18 10 10.7 & -- & -- &
153$\pm$1 & 2 & 75 & 0.5 / 8 & 12 \\
CI Tau & 04 33 51.99 & 22 50 30.6 & -- & -- &
40$\pm$18 & 8 & 54 & 1.0 / 4 & 76 \\
AA Tau & 04 34 55.40 & 24 28 53.8 & -- & -- &
75$\pm$16 & 6 & 51 & 0.5 / 21 & 65 \\
DN Tau & 04 35 27.35 & 24 14 59.8 & -- & -- &
120$\pm$5 & 6 & 52 & 0.5 / 22 & 22 \\
LkCa 15 & 04 39 17.80 & 22 21 04.5 & -- & -- &
61$\pm$1 & 2 & 54 & 1.5 / 4 & 83 \\
GM Aur & 04 55 10.95 & 30 22 01.0 & -- & -- &
51$\pm$2 & 2 & 65 & 2.0 / 9 & 76 \\
\hline
\end{tabular}
\label{tab:datatts}

References: 1) Dougados et al. (2000); 2) Simon et al. (2000);
3) Gomez de Castro \& Pudritz (1992); 4) Padgett et
al. (1999); 5) M\'enard et al. (2004); 6) Kitamura et
al. (2002); 7) Mundt et al. (1991); 8) Dutrey et al. (1996);
9) Lavalley et al. (1997); 10) Kitamura, Kawabe \& Saito
(1996); 11) Eisl\"offel \& Mundt (1998); 12) Hirth et
al. (1997); 13) Stapelfeldt et al. (1998); 14) Stapelfeldt et
al. (2004, in prep.); 15) Monin \& Bouvier (2000); 16) Mundt
\& Eisl\"offel (1998); 17) Close et al. (1998); 18) Solf \&
B\"ohm (1999); 19) Akeson, Koerner \& Jensen (1998); 20)
Wilner, Ho \& Rodriguez (1996); 21) Mundt et al. (1990); 22)
Guilloteau, Dutrey \& Simon (1999); 23) Burrows et al. (1996);
24) Aspin \& Reipurth (2000); 25) Strom et al. (1986); 26)
Hartigan et al. (2004).
\end{table*}

When available, we have quoted published error bars for the
orientations.  The following estimates may be
used as guidelines.  Collimated jets are usually identified in
deep narrow-band images (e.g., Mundt, Ray \& Raga 1991;
Dougados et al. 2000) or through long-slit (e.g., Hirth, Mundt
\& Solf 1997) or slitless (Hartigan, Edwards \& Pierson 2004)
spectroscopic observations. In most cases, jets are clearly
resolved and their PAs are known within $\pm$10\degr\ or
better.

Disks around young stars can be identified in two main ways: thermal
imaging in the submillimeter and millimeter ranges and scattered light
imaging in the optical and near-infrared. For practical reasons, the
latter technique is usually limited to edge-on disks (e.g., Burrows et
al. 1996; Stapelfeldt et al. 1998) and to circumbinary disks (e.g.,
Roddier et al. 1996; Close et al. 1998). For the former, a dark lane
is usually well defined and the PAs are known very well, usually
$\pm1$\degr\ . This is the case for HH~30, HV~Tau~C, and HK~Tau~B for
example.

At radio wavelengths, dust continuum images of disks are
obtained with long baseline interferometers (e.g., Dutrey et
al. 1996; Kitamura et al. 2002). Furthermore, molecular-line
images of disks reveal clear velocity gradients that are
consistent with Keplerian rotation (e.g., Simon, Dutrey \&
Guilloteau 2000). When available, we used the resolved CO maps
to define the orientation of the disk's semi-major axis. Disk
orientations are known to $\pm$5\degr\ or so when a Keplerian
velocity gradient is detected and to $\pm$5--15\degr\
otherwise, depending on the inclination.

\subsubsection{Notes on individual objects}
\label{sec:remarks}
\paragraph{\sl DD Tau.}
Gomez de Castro \& Pudritz (1992) observed a double-peaked
structure for DD~Tau which they interpreted as ``focal
points'' of a jet above and below the disk plane. However,
Leinert et al. (1993) found DD~Tau to be a binary system with
the same separation and PA. Furthermore, Hartigan
\& Kenyon (2003) obtained visible spectra of each component.
They both show a continuum, with an M3.5 spectral type,
rejecting the possibility that these peaks are stationary
shocks in a jet. On the other hand, Hartigan et al. (2004)
obtained slitless spectroscopy images of the system, revealing
a jet emanating from the primary at PA $\sim125$\degr, along
which Gomez de Castro \& Pudritz (1992) had indeed observed
spatially-resolved low level forbidden line emission. We adopt
this PA for the jet of DD~Tau~A in this study.
\paragraph{\sl FT~Tau.}
Stapelfeldt et al. (2004, in prep.) suggest the presence of a small
HH-like nebulosity at 1\farcs0 and PA = 209\degr\ from FT~Tau. This is
the value quoted in Table~\ref{tab:datatts} for the jet PA, once
transformed to the range [0,180[. The feature is at the detection
limit of their HST/WFPC2 image, and remains unconfirmed. The PA given
by Dutrey et al. (1996), 82\degr$\pm$17\degr, for the dust continuum
emission detected at 2.7mm, i.e., the PA of the disk, relies on data
obtained with a low resolution configuration at the IRAM
interferometer.  The PA of the jet is better defined and we use this
value in the comparison with the magnetic field.

\paragraph{\sl UZ~Tau~E.}
The CO~disk suggested by Jensen et al. (1996) is aligned very
closely with the optical jet detected by Hirth et al. (1997)
and roughly perpendicular to the Keplerian structure clearly
identified by Simon et al. (2000). Since the latter
observations have a higher spatial resolution and better
sensitivity, we assume that the structure discovered by Jensen
et al. corresponds to an outflow instead of a disk. As a
confirmation, the optical jet has recently been imaged by
Hartigan et al. (2004).
\paragraph{\sl DO~Tau.} 
The extended millimeter-wave structure around DO~Tau identified by
Kitamura et al. (2002) is oriented along the direction of the
optical jet identified by Hirth et al. (1997). It is likely
that the emission they detect is in fact related to the jet
(e.g., free-free emission) rather than being a disk-like
structure. In the following, we consider only the orientation
estimated from the optical jet.
\paragraph{\sl Other sources.}
Finally, we note that the disks surrounding UZ~Tau~E, UY~Aur and
GG~Tau are not circumstellar disks, but circumbinary disks or
rings. The jets associated with the first two systems are likely to come
from one of the two stars and it is unclear whether the orientations of
the circumbinary disks in these systems are related to the inner binary
orientation.  However, for both UZ~Tau~E and UY~Aur, the detected jets
appear to be perpendicular to the circumbinary disks. For lack of
better information regarding the inner circumstellar disks in these
binary systems, we assume that they are coplanar and keep the PAs of
the circumbinary disks in the analysis below.

\subsection{Direction of the magnetic field in Taurus}

We use published linear polarisation measurements of
background stars to trace the direction of the magnetic field
projected in the plane of the sky in Taurus. We compiled the
polarisation measurements published by Vrba, Strom \& Strom
(1976), Heyer et al. (1987), Moneti et al. (1984), Tamura et
al. (1987), Tamura \& Sato (1989), and Goodman et al. (1990,
1992). These studies cover most of the area we are interested
in. To complete the polarisation data set in the vicinity of
T~Tau itself, we used the polarisation vectors shown in
Figure~1c of Tamura \& Sato (1989). Similarly, in the areas
located at $\alpha_{\rm 2000}=$ 04h05m and $\delta_{\rm
2000}=$ +26\degr\ and 04h15m$\leq \alpha_{\rm 2000} \leq
$04h25m and +28\degr $\leq \delta_{\rm 2000} \leq$+30\degr ,
we estimated the local polarisation from Figure~2 of Tamura \&
Sato (1989). Overall, we compiled $\sim$400 polarisation
measurements.  Only measurements with $P/\sigma_P > 3$ were
retained to ensure a good determination of the PA,
and hence of the projected direction of the magnetic field.

%------figure 1-------------
\begin{figure*}[t]
\centering
\includegraphics[height=11cm,width=11.5cm]{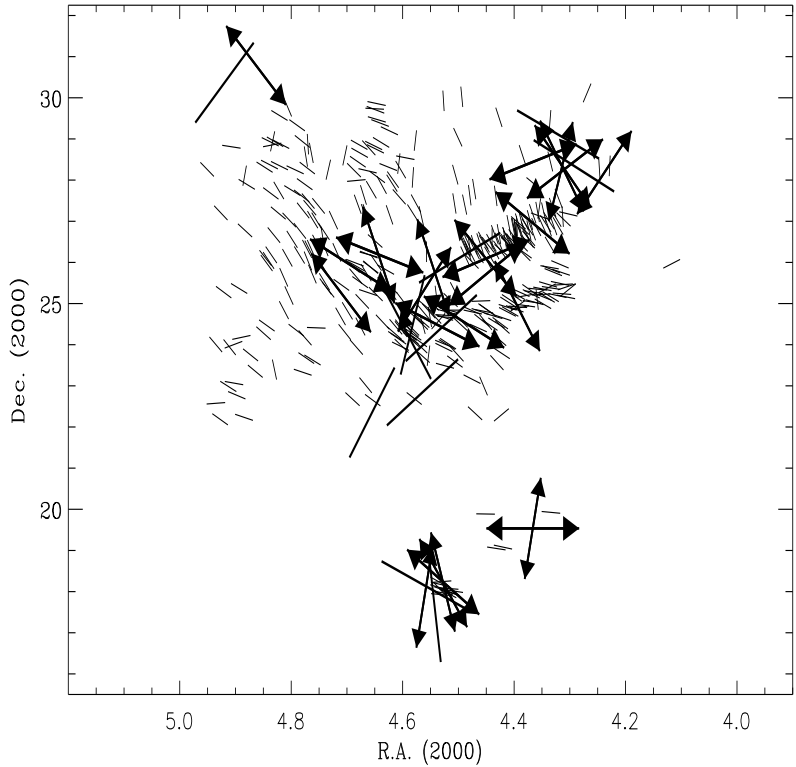}
\caption{Plot of the orientation of the stars with respect to the 
  magnetic field. Small segments represent interstellar polarisation
  measurements while long bold segments indicate the orientation of
  CTTS. Thick vectors with arrow heads are for CTTS with jets, thick
  vectors without arrow heads are for CTTS with a disk but no detected
  extended outflow. Short thin segments trace the direction of the
  magnetic field.}
\label{fig:taurusall}
\end{figure*}
%-------------------------

To estimate the direction of the magnetic field near each
target T~Tauri star, we searched the interstellar polarisation
database in concentric circles starting with radii of 0\fdg5,
and increasing to 1\fdg0, 1\fdg5, and 2\fdg0. We selected the
smallest radius containing at least 4 different
measurements. The median of the PAs in this circle
is used as the direction of the local magnetic field close to
each T~Tauri stars.

UY~Aur and GM~Aur are located outside the zone containing the
bulk of polarisation measurements. In these two cases, the
direction of the magnetic field is extrapolated from the
nearest measurements, as opposed to be medianed in an area
centered on the sources. Within 2\degr\ or so of both sources
the magnetic field is well ordered and the extrapolation
should be reliable.

\section{Results} 

In Table~\ref{tab:datatts} we present the PAs we
compiled for CTTS. All angles are measured East of North and
are given in the range [0\degr, 180\degr [. Columns~1, 2, and
3 contain the object name and coordinates, respectively. The
PA of the jet, P.A.~(jet), and the reference are
given in Cols.~4 and 5. The PA of the major
axis of the disk, P.A.~(disk), and the reference are in
Cols.~6, and 7. Col.~8 is the estimation of the projected
direction of the magnetic field, P.A.~({\bf B}), in the vicinity of
the source. For the interstellar polarisation measurements,
$\rho_{IS}$ and $N_{IS}$ are the radius within which the
orientation of the magnetic field is estimated and the number
of measurements considered, respectively. They are shown in
Col.~9. Finally, the last column contains
$|\Delta$P.A.$|$, the difference in PA between the
local magnetic field and the symmetry axis of the CTTS. This
is the quantity we are interested in.

The results are plotted in Figure~\ref{fig:taurusall}. The
short, thin vectors trace the projected direction of the
magnetic field across the cloud as measured by linear
polarisation of background stars. The size of the vectors is
uniform and not to scale with the level of linear
polarisation. The large scale magnetic field runs roughly in
the NE-SW direction (e.g., Tamura \& Sato 1989), with
significant bending on smaller scales.

The long, thick vectors trace the projected orientation of
each CTTS. Thick vectors with arrow-heads are for sources with
a resolved jet/outflow and trace the direction of the
flow. Thick vectors without arrow-heads are for sources where
only a disk is known today, they trace the direction
perpendicular to the major-axis of the disk.  Therefore, all
thick vectors are tracing the same symmetry axis for the CTTS,
namely the axis of rotation (i.e., along the jet or
perpendicular to the disk, which is similar). It is also
important to note that the error bars on the orientation of
CTTS, typically 5\degr--10\degr\, are not reponsible for the
large spread observed. The wide distribution of PAs with
respect to the magnetic field is a property of the CTTS
sample.

\section{Orientation of CTTS in Taurus-Auriga}
\label{sec:random}

Throughout the discussion below we will assume that the
symmetry axis of a CTTS+disk system is the same as its
rotation axis and its jet/outflow axis. This assumption
reflects our current observational and theoretical
understanding of the geometry of a T~Tauri star and its
circumstellar environment. In all figures and histograms, only
the symmetry axis of the systems is used. In other words, the
90\degr\ difference between the disk major axis and the jet
axis has been taken into account in all plots.

\begin{figure}[t]
\centering
\includegraphics[width=\columnwidth]{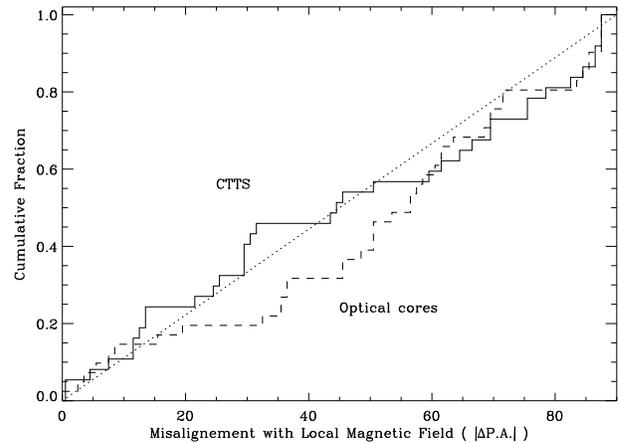}
\caption{Cumulative distribution function of the difference in
  PAs between the local magnetic field and the
  CTTS symmetry axis (solid histogram,
  Table\,\ref{tab:datatts}) and the major axis of optical
  cores (dashed histogram, from Lee \& Myers 1999). The dotted
  line is the function expected for an infinite randomly oriented
  sample.}
\label{fig:cdf_ctts_cores}
\end{figure}

In Figure~\ref{fig:cdf_ctts_cores} we present the cumulative
distribution histogram (CDH) of the CTTS sample. The straight
dotted-line is the CDH expected for an infinite perfectly random
sample. The solid histogram of the complete CTTS sample follows this
line closely. Statistically, the null hypothesis from a Kolmogorov
test is accepted, i.e., the observed distribution of CTTS can be drawn
from a randomly oriented parent distribution, with only 20\% chance
to reject the null hypothesis. This result indicates that {\it the
group of CTTS we compiled is very likely randomly oriented with
respect to the local magnetic field in Taurus}.

This new result contradicts previous claims, e.g., by Strom et
al.~(1986) and Tamura \& Sato~(1989), that young stars were
aligned mostly parallel to the magnetic field. The origin of
that contradiction lies in the larger size of our sample and
in biases in the composition of previous samples.

%----------------------
%---------------------

In the study of Strom et al. (1986) the sample was incomplete
and made only of stars with well defined jets or bright
Herbig-Haro objects. Many of their sources are included in
Table~\ref{tab:datatts}; those not included are not CTTS but
more embedded YSOs. In the study of Tamura \& Sato (1989), the
orientation of the YSOs was deduced from near-infrared
polarimetry. This is a more indirect method that relies on
models. Caution must also be used as it furthermore suffers from a
90\degr\ ambiguity, with polarisation vectors being parallel
or perpendicular to the disk depending whether or not an
envelope is present (e.g., Whitney \& Hartmann 1993). This
ambiguity can not be lifted easily without direct
imaging. Therefore, the samples used in previous studies were
biased towards sources driving bright jets, or with a poorly
or ambiguously defined geometry.

On the opposite, the sample in Table~\ref{tab:datatts} is made
only of stars where the disks and/or jets are well
resolved. The orientation of all sources projected in the
plane of the sky is secure and determined without
ambiguity. We note that there are selection effects against
the detection of both jets and disks in the case of pole-on
systems. However, it is unlikely that this bias prevents us
from including targets at specific PA with respect
to the magnetic field {\it projected on the plane of the
sky}. Furthermore, our sample contains CTTS only, removing the
possible confusion arising from the presence of an extended
and dusty envelope around the class I sources contained in
other samples.

\section{The influence of magnetic field throughout star
formation}

On the large scale, star formation in the Taurus-Auriga
molecular cloud appears to be at least partially driven by the
magnetic field. First, both the dense gas clouds and the YSOs
show a large scale filamentary distribution roughly
perpendicular to the magnetic field. Furthermore, at the
smaller, individual object scale, Hartmann (2002) showed that
pre-stellar cores, which are most likely prolate (Curry 2002),
are elongated preferentially parallel to the direction of the
filaments. These findings contrast with our main result that
individual CTTS are randomly oriented with the local magnetic
field at a typical age of 1-3~Myr. Are these apparently
contradicting results revealing a physical mechanism at play
during the late stages of star formation or the mere
consequence of poor statistics associated with small samples?

In Figure\,\ref{fig:cdf_ctts_cores}, we have also plotted the CDH of
the relative angle between the major axis of optical cores from Lee \&
Myers (1999) and the local magnetic field in Taurus. We used the
orientation of the long axis of the cores as their symmetry axis since
they are likely to be prolate (Curry 2002). We have included all cores
in the same area of the sky where we have studied CTTS and have used
the same method to determine the orientation of the local magnetic
field. The CDH of optical cores is systematically under the
theoretical distribution for random orientation and the median
relative angle is about 55\degr. As pointed out by Hartmann (2002),
the alignment of cores is uneven across Taurus. For example, it is
particularly good in a few regions, e.g., his group 3. In general,
optical cores have their major axes preferentially oriented
perpendicularly to the local magnetic field. Therefore, as a group,
the probability (from a Kolmogorov test) that their distribution can
be drawn from a randomly oriented parent distribution is small, only
$\sim$8\,\%.

Similarly, the probability that the samples of CTTS and optical cores
can be taken from the same randomly oriented parent distribution is
only about 11\,\%.  In other words, the moderate trend observed for
optical cores to be preferentially oriented perpendicular to the local
magnetic field does not seem to apply anymore at the later CTTS stage.

The available observations of the Taurus-Auriga molecular
cloud and its populations of prestellar cores and CTTS
therefore suggest a scenario in which {\it the role of the
cloud's magnetic field decreases as star formation proceeds to
ever smaller scales}. As suggested by Hartmann (2002), the
magnetic field has likely driven the early collapse of the
entire cloud into several regularly-spaced filaments that are
perpendicular to the large scale magnetic field. On the
intermediate scale, that of individual cores, this preferred
orientation of the cores is still observed, though as a weaker
trend. At small scales however, moving towards the formation
of individual star+disk systems out of dense cores, the memory
of the original direction of collapse appears lost as
indicated by the observed random orientation of the CTTS'
symmetry axis in the same reference frame of the magnetic
field.

This evolution of the distribution of PAs from large to small scales
does not necessarily invalidate the current paradigm of isolated,
quasi-static star formation. However, it indicates that during the end
stages of the star formation process the final orientation of a system
may be determined by a stochastic mechanism that becomes independent
of the magnetic field.

This decoupling from the early stages could come from the fact
that prolate prestellar cores may first evolve towards a
quasi-spherical configuration due to their self-gravity (Curry
2002), in line with the observation that protostellar cores
are rounder than starless cores (Goodwin, Ward-Thomson \&
Whitworth 2002). Alternatively or additionally, these cores
could fragment into two or more smaller and more spherical
cores, as needed to account for the very high binary frequency
among young stars (Hartmann 2002). Because of their proximity,
these rounder cores would be more prone to dynamical
interactions (for example) that could randomize their
orientations.

\section{Exploring the influence of orientation on the jet 
properties}
\label{sec:spec}

In this section we explore the impact of the orientation of a
CTTS with respect to the large scale magnetic field on its
capacity to launch a powerful and a well collimated
outflow. Surprisingly, while a connection between the (random)
orientation of a T Tauri stars and the strength of its outflow
is a priori not expected in the current models, our results
may indicate othwerwise.

\begin{figure}[t]
\centering
\includegraphics[width=\columnwidth]{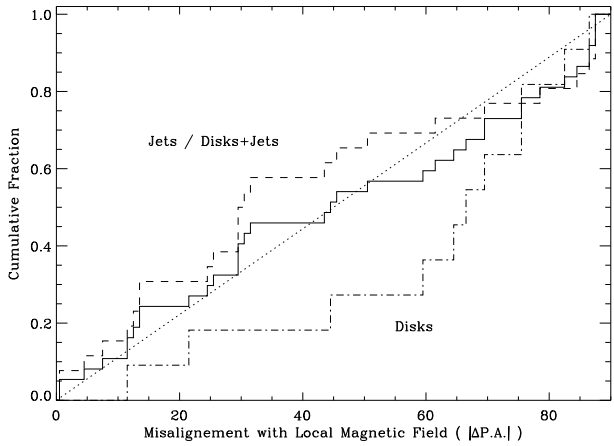}
\caption{Cumulative distribution function of the difference in
  PAs between the local magnetic field and the
  CTTS symmetry axis (Col.10 in Table\ref{tab:datatts}). The
  dashed histogram is for all sources with a jet, the
  dot-dashed histogram for sources with a disk but no
  jet/outflow and the solid histogram is for the whole
  sample. The dotted line is the function expected for an infinite
  randomly oriented sample.}
\label{fig:cdf_jets_disks}
\end{figure}

\begin{figure}[t]
\centering
\includegraphics[width=\columnwidth]{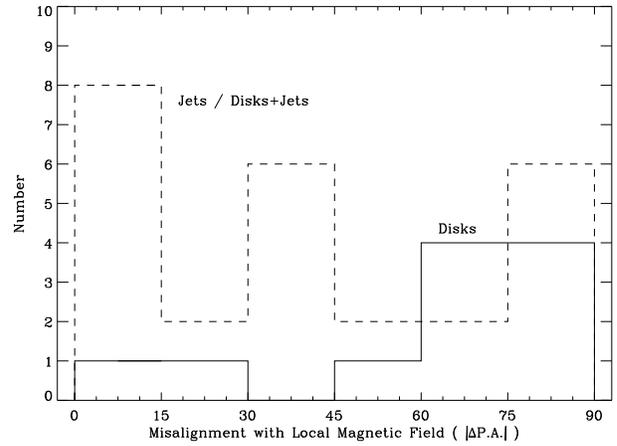}
\caption{Histogram of the difference in PAs between
  the objects' symmetry axes and the local magnetic field. The
  solid-line histogram represent CTTS with disks but no jet,
  the dashed-line histogram is for CTTS with a jet.}
\label{fig:histo}
\end{figure}

Consider all CTTS where only a disk is detected, i.e., the bottom part
of Table~\ref{tab:datatts}. Surprisingly, they appear to align
preferentially at a large angle from the local magnetic field (i.e.,
with the major-axis of their disk {\it parallel} to the magnetic
field). This is illustrated in both Figure~\ref{fig:cdf_jets_disks}
and Figure~\ref{fig:histo}. The probability that the CDH for this
subsample is drawn from a randomly distributed sample or from the same
parent distribution as the CTTS with a resolved jet are 5\,\% and
13\,\%, respectively. Due to the small number of sources with only a
disk in our sample, these levels are not conclusive. However, they
suggest that objects for which a jet has not yet been detected are
oriented {\it differently} than the complete sample, with a preference
to be perpendicular to the local magnetic field.

It is interesting to note however that almost all these
sources also have forbidden emission lines which trace
mass-loss activity in their spectrum (Cabrit et al. 1990;
Hartigan, Edwards \& Ghandour 1995; Hartigan \& Kenyon
2003). This means that mass-loss is most likely present in all
the CTTS of Table~\ref{tab:datatts}, but it is either not
extended enough or it is too weak to be resolved for the
so-called disk-only sources.

Previous claims regarding the orientation of CTTS were based on
subsamples made of sources driving bright and well collimated flows.
They showed a tendency to align parallel to the magnetic field (e.g.,
Strom et al. 1986).  With improved imaging techniques, weaker and/or
shorter jets are now being discovered (e.g., Hartigan et al. 2004;
Stapelfeldt et al. 2004, in prep.)  Surprisingly, they are found to
originate more often from objects whose symmetry axis is oriented far
away from the magnetic field. These new jets destroy the correlation
previously found.

This lack of correlation is interesting since all CTTS probably still
have mass-loss based on the presence of forbidden line emission in
their spectrum. As a consequence, the correlation identified in early
studies likely does not exist, it is the result of small sample size
and selection effects\footnote{This is a refinement of our previous
suggestion that a better alignment with the magnetic field was found
for this sample (Duch\^ene, \& M\'enard 2003; M\'enard \& Duch\^ene
2004). At that time, the subsample of sources with spatially resolved
jets did not include recent detections of very low surface brightness,
poorly collimated outflows.}.

An interesting possibility to interpret this result is to
suggest that {\it there is a connection between the
orientation of a T Tauri star and the strength and/or length
of its jet}, namely that systems roughly aligned with the
local magnetic field are more prone to drive bright, extended
and well-defined jets while those largely misaligned cannot
develop the proper conditions to drive extensive jet.

%----------------------
%---------------------

This interpretation must still be confirmed with a more
quantitative analysis of the jet properties (e.g., surface
brightness in specific emission lines, spatial extent,
collimation angle, etc.). Yet, our results support such a
possibility. If confirmed by future studies, it could indicate
that somehow there is a feedback between the configuration of
the magnetic field at the stellar surface / inner disk
and the direction of the nearby field in the molecular cloud.

Current theories of ejection mechanisms for CTTS are based either on
the presence of a relatively strong {\it stellar} magnetic field (Shu
et al. 2000) whose long-range dipolar component interacts with the
inner disk or based on the presence of a non-stellar field threading
the inner disk (K\"onigl \& Pudritz 2000; Ferreira 1997). The size and
topology of the jet-launching region, e.g., more or less extended,
depends on the exact model. Noticeably however, Ferreira (1997) showed
in the framework of an MHD disk-wind model that a dominant quadrupolar
magnetic field configuration leads to a much weaker wind than a
dipolar configuration. Our findings therefore support the speculative
idea that the magnetic field configuration in the disk and/or
magnetosphere of a CTTS may be influenced by its relative orientation
with respect to the cloud's magnetic field.

%
%----------------------------------------
\section{Summary and future work}

We used recent high-angular resolution images of young stars
to re-evaluate the importance of the local magnetic field in
the late stages of the star formation process. We have
focussed on the quiescent Taurus-Auriga star-forming region
where massive stars and their strong influence on the cloud
dynamics are absent. We have compiled a database of 37 CTTS
with spatially resolved collimated jets and/or circumstellar
disks. For each object, we have determined the orientation of
its symmetry axis on the plane of the sky and compared it to
the direction of the local magnetic field.

Whereas previous studies had suggested that young stars form with
their symmetry axis parallel to the magnetic field, {\sl we find that
the population of CTTS in Taurus-Auriga is randomly oriented with
respect to the magnetic field}. This is the main result of this
work. It suggests either that the influence of the cloud's magnetic
field on the formation process is dominant at large scales (entire
cloud) but largely decreases on the much smaller scale of individual
objects or that the orientation has changed since birth.

We also find a speculative connection between the strength of
CTTS jets and their orientation with respect to the magnetic
field. Bright, elongated well-collimated jets are
preferentially parallel to the magnetic field while weaker,
fuzzy, and/or shorter jets tend to be perpendicular to
it. This may indicate a link between the orientation of an
object with respect to the cloud's magnetic field and the
morphology (e.g., dipolar vs. quadrupolar) of the stellar
magnetic field it can sustain at long range.

Deeper imaging of CTTS will reveal more jets and disks,
allowing to test this suggestion. In particular, it will be
interesting to determine the properties (surface brightness,
length, collimation angle) of the jets from those sources for
which only a disk has been resolved so far. One can also
imagine using submillimeter linear polarisation measurements
to study the orientation of much younger embedded protostars.
Finally, conducting
similar studies in other star-forming regions, such as the
much denser $\rho$\,Ophiuchus and Orion clouds, will help
disentangle the influence of magnetic field on star formation
from other physical effects.

\begin{acknowledgements}

The authors acknowledge interesting discussions with Catherine
Dougados and Jonathan Ferreira. FM also wishes to thank the
organisers of the workshop on {\it Solar and Stellar
Magnetism} held in Toulouse, France, in September 2002 where
we had the idea presented in this paper. Financial support
from the ``Programme National de Physique Stellaire'' (PNPS)
of CNRS/INSU, France, is gratefully acknowledged.  This
research has made use of the SIMBAD database, operated at CDS,
Strasbourg, France.

\end{acknowledgements}

\end{document}